\documentclass[aps,prl,superscriptaddress,showpacs,preprintnumbers,amsmath,amssymb]{revtex4}

\usepackage{graphicx} 
\usepackage{dcolumn}  
\usepackage{subfigure}

\newcommand{\ee}{\mbox{\ensuremath{e^+ e^-}}}

\newcommand{\RM}{\ensuremath{M_{\mathrm{recoil}}}}

\newcommand{\gevc}{\ensuremath{\, {\mathrm{GeV}/c^2}}}
\newcommand{\mevc}{\ensuremath{\, {\mathrm{MeV}/c^2}}}

\newcommand{\mev}{\ensuremath{\, {\mathrm{MeV}}}}

\newcommand{\ifb}{\ensuremath{\, {\mathrm{fb^{-1}}}}}

\newcommand{\el}{\mbox{\ensuremath{\ell^+ \ell^-}}}

\newcommand{\ups}{\mbox{\ensuremath{\Upsilon(4S)}}}
\newcommand{\jp} {\mbox{\ensuremath{J/\psi}}}
\newcommand{\pp} {\mbox{\ensuremath{\psi(2S)}}}
\newcommand{\et} {\mbox{\ensuremath{\eta_c}}}
\newcommand{\etp}{\mbox{\ensuremath{\eta_c(2S)}}}

\newcommand{\pin} {\mbox{\ensuremath{\pi^0}}}
\newcommand{\ks} {\mbox{\ensuremath{K^0_S}}}

\newcommand{\drec}{\mbox{\ensuremath{D_{\text{rec}}}}}
\newcommand{\dtag}{\mbox{\ensuremath{\overline{D}{}_{\text{assoc}}}}}
\newcommand{\darec}{\mbox{\ensuremath{D^{(*)}_{\text{rec}}}}}
\newcommand{\datag}{\mbox{\ensuremath{\overline{D}{}^{(*)}_{\text{assoc}}}}}
\newcommand{\dsrec}{\mbox{\ensuremath{D{}^{*}_{\text{rec}}}}}
\newcommand{\dstag}{\mbox{\ensuremath{\overline{D}{}^{*}_{\text{assoc}}}}}

\newcommand{\dd}{\mbox{\ensuremath{D \overline{D}}}}
\newcommand{\dds}{\mbox{\ensuremath{D^* \overline{D}}}}
\newcommand{\dsds}{\mbox{\ensuremath{D^* \overline{D}{}^*}}}
\newcommand{\da}{\mbox{\ensuremath{D^{(*)}}}}
\newcommand{\dab}{\mbox{\ensuremath{ \overline{D}{}^{(*)}}}}
\newcommand{\dda}{\mbox{\ensuremath{D^{(*)} \overline{D}{}^{(*)}}}}

\newcommand{\eedd}{\mbox{\ensuremath{\ee \! \to \! \jp \dd }}}
\newcommand{\eedds}{\mbox{\ensuremath{\ee \! \to \! \jp \dds}}}
\newcommand{\eedsds}{\mbox{\ensuremath{\ee \! \to \! \jp \dsds}}}
\newcommand{\eedda}{\mbox{\ensuremath{\ee \! \to \! \jp\ \dda}}}

\def\bbr{{\mbox{\sl B\hspace{-0.4em} {\footnotesize\sl
A}\hspace{-0.4em} B\hspace{-0.4em} {\footnotesize\sl
A\hspace{-0.1em}R}}}}

\begin{document}
\title{ \quad\\[0.5cm] \Large Production of new charmonium-like states in
$\mathbf {\ee \! \to \! \jp \dda}$ at $\mathbf{\sqrt{s} \approx
10.6}\,$GeV}


\affiliation{Budker Institute of Nuclear Physics, Novosibirsk}
\affiliation{Chiba University, Chiba}
\affiliation{University of Cincinnati, Cincinnati, Ohio 45221}
\affiliation{Justus-Liebig-Universit\"at Gie\ss{}en, Gie\ss{}en}
\affiliation{The Graduate University for Advanced Studies, Hayama}
\affiliation{Hanyang University, Seoul}
\affiliation{University of Hawaii, Honolulu, Hawaii 96822}
\affiliation{High Energy Accelerator Research Organization (KEK), Tsukuba}
\affiliation{University of Illinois at Urbana-Champaign, Urbana, Illinois 61801}
\affiliation{Institute of High Energy Physics, Chinese Academy of Sciences, Beijing}
\affiliation{Institute of High Energy Physics, Vienna}
\affiliation{Institute of High Energy Physics, Protvino}
\affiliation{Institute for Theoretical and Experimental Physics, Moscow}
\affiliation{J. Stefan Institute, Ljubljana}
\affiliation{Kanagawa University, Yokohama}
\affiliation{Korea University, Seoul}
\affiliation{Kyungpook National University, Taegu}
\affiliation{\'Ecole Polytechnique F\'ed\'erale de Lausanne (EPFL), Lausanne}
\affiliation{University of Ljubljana, Ljubljana}
\affiliation{University of Maribor, Maribor}
\affiliation{University of Melbourne, School of Physics, Victoria 3010}
\affiliation{Nagoya University, Nagoya}
\affiliation{Nara Women's University, Nara}
\affiliation{National Central University, Chung-li}
\affiliation{National United University, Miao Li}
\affiliation{Department of Physics, National Taiwan University, Taipei}
\affiliation{H. Niewodniczanski Institute of Nuclear Physics, Krakow}
\affiliation{Nippon Dental University, Niigata}
\affiliation{Niigata University, Niigata}
\affiliation{University of Nova Gorica, Nova Gorica}
\affiliation{Osaka City University, Osaka}
\affiliation{Osaka University, Osaka}
\affiliation{Panjab University, Chandigarh}
\affiliation{RIKEN BNL Research Center, Upton, New York 11973}
\affiliation{Saga University, Saga}
\affiliation{University of Science and Technology of China, Hefei}
\affiliation{Seoul National University, Seoul}
\affiliation{Sungkyunkwan University, Suwon}
\affiliation{University of Sydney, Sydney, New South Wales}
\affiliation{Tata Institute of Fundamental Research, Mumbai}
\affiliation{Toho University, Funabashi}
\affiliation{Tohoku Gakuin University, Tagajo}
\affiliation{Department of Physics, University of Tokyo, Tokyo}
\affiliation{Tokyo Institute of Technology, Tokyo}
\affiliation{Tokyo Metropolitan University, Tokyo}
\affiliation{Tokyo University of Agriculture and Technology, Tokyo}
\affiliation{Virginia Polytechnic Institute and State University, Blacksburg, Virginia 24061}
\affiliation{Yonsei University, Seoul}
   \author{I.~Adachi}\affiliation{High Energy Accelerator Research Organization (KEK), Tsukuba} 
   \author{H.~Aihara}\affiliation{Department of Physics, University of Tokyo, Tokyo} 
   \author{K.~Arinstein}\affiliation{Budker Institute of Nuclear Physics, Novosibirsk} 
   \author{T.~Aushev}\affiliation{\'Ecole Polytechnique F\'ed\'erale de Lausanne (EPFL), Lausanne}\affiliation{Institute for Theoretical and Experimental Physics, Moscow} 
   \author{T.~Aziz}\affiliation{Tata Institute of Fundamental Research, Mumbai} 
   \author{A.~M.~Bakich}\affiliation{University of Sydney, Sydney, New South Wales} 
   \author{V.~Balagura}\affiliation{Institute for Theoretical and Experimental Physics, Moscow} 
   \author{E.~Barberio}\affiliation{University of Melbourne, School of Physics, Victoria 3010} 
   \author{I.~Bedny}\affiliation{Budker Institute of Nuclear Physics, Novosibirsk} 
   \author{K.~Belous}\affiliation{Institute of High Energy Physics, Protvino} 
% ...
\collaboration{The Belle Collaboration}


\begin{abstract}
We present a study of the $X(3940)$ state in the process $\ee \to \jp
\,D^{*} \overline{D}$. The $X(3940)$ mass and width are measured to be
$(3942\,^{+\,7}_{-\,6} \pm 6)\mevc$ and
$\Gamma=(37\,^{+\,26}_{-\,15}\pm 8)$\mev. In the process $\ee \to \jp
\,D^{*+} D^{*-}$ we have observed another charmonium-like state, which
we denote as $X(4160)$, in the spectrum of invariant masses of $D^{*+}
D^{*-}$ combinations. The $X(4160)$ parameters are $M=
(4156\,^{+\,25}_{-\,20} \pm 15)\mevc$ and $\Gamma =
(139\,^{+\,111}_{-\,\phantom{1}61} \pm 21)\mev$. The analysis is based
on a data sample with an integrated luminosity of $693\ifb$ recorded
near the \ups\ resonance with the Belle detector at the KEKB
\ee\ asymmetric-energy collider.
\end{abstract}

\maketitle
\setcounter{footnote}{0}

\noindent Double charmonium production in \ee\ annihilation, first
observed by Belle in 2002~\cite{2cc}, can be used to search for new
charmonium states, recoiling against some known and easily
reconstructed charmonium. The study of various double charmonium
final states~\cite{2cc2,babar_2cc} demonstrated that there is no
significant suppression of the production of radially excited states:
the cross-sections for \jp\et, \pp\et, \jp\etp\ and \pp\etp\ are very
close to each other. These studies also show that scalar and
pseudoscalar charmonia are produced copiously recoiling against \jp\
or \pp. A new charmonium-like state, $X(3940)$, has been already
observed in the spectrum recoiling against \jp, and reconstructed in
the \dds~\cite{cc} final state~\cite{x3940}. On the other hand, there
has recently been a number of reports on observation of new
charmonium or charmonium-like states above \dd\
threshold~\cite{new_cc}. Their properties are quite different from
those expected from the quark model. These experimental results have
renewed theoretical interest in spectroscopy, decay and production of
charmonia~\cite{theor}.

In this Letter we present a new study of the $X(3940)$ resonance and
report on the observation of a new charmonium-like state in the
process \eedda\ and the measurement of its parameters. The integrated
luminosity used for this analysis is $693\ifb$ collected with the
Belle detector~\cite{Belle} near the \ups\ resonance at the KEKB
asymmetric-energy \ee\ collider~\cite{KEKB}.


This study is performed using the selection procedure similar to that
described in Ref.~\cite{2cc,x3940}. All charged tracks are required to
be consistent with originating from the interaction point.  Charged
kaon candidates are required to be positively identified, while no
identification requirements are applied for pion candidates as the
pion multiplicity is much higher than those of other
hadrons. \ks\ candidates are reconstructed by combining $\pi^+ \pi^-$
pairs with an invariant mass within $10\,\mathrm{MeV}/c^2$ of the
nominal \ks\ mass. We require the distance between the pion tracks at
the \ks\ vertex to be less than $1\,\mathrm{cm}$, the transverse
flight distance from the interaction point to be greater than
$1\,\mathrm{mm}$ and the angle between the \ks\ momentum direction and
decay path to be smaller than $0.1\,\mathrm{rad}$. Photons are
reconstructed in the electromagnetic calorimeter as showers with an
energy more than $20 \mev$ that are not associated with charged
tracks. Photons of energy more than $50 \mev$ are combined to form
\pin\ candidates. If the mass of $\gamma \gamma$ pairs lies within $15
\mevc$ of the nominal \pin\ mass, such pairs are fitted with a
\pin\ mass constraint and considered as
\pin\ candidates. \jp\ candidates are reconstructed via the $\jp \!
\to \!  \el$ ($\ell=e, ~\mu $) decay channel.  Two positively
identified lepton candidates are required to form a common vertex that
is less than $1\,\mathrm{mm}$ ($\approx 6\,\sigma$) from the
interaction point in the plane perpendicular to the beam axis. A
partial correction for final state radiation and bremsstrahlung energy
loss is performed by including the four-momentum of every photon
detected within a $50\,\mathrm{mrad}$ cone around the electron
direction in the \ee\ invariant mass calculation. The \jp\ signal
region is defined by the mass window $\left|M_{\ell^{+} \ell^{-}} -
M_{J/\psi}\right| \! < \! 30 \mevc$ ($\approx \! 2.5\,
\sigma$). \jp\ candidates are subjected to a mass-vertex fit to
improve their momentum resolution. QED processes are suppressed by
requiring the total charged multiplicity in the event to be more than
4. \jp\ mesons from $B\overline{B}$ events are removed by requiring a
center-of-mass (CM) momentum $p^{*}_{J/\psi}>2.0\,\mathrm{GeV}/c$.

We reconstruct $D^0$ mesons using five decay modes: $K^- \pi^+$, $K^-
K^+$, $K^- \pi^- \pi^+ \pi^+$, $\ks \pi^+ \pi^-$ and $K^- \pi^+
\pin$. Candidate $D^+$ mesons are reconstructed using $K^- \pi^+
\pi^+$, $K^- K^+ \pi^+$ and $\ks \pi^+$ decay modes. A $\pm 15\mevc$
mass window is used for all modes except $D^0 \to K^- \pi^+ \pi^0$
($\pm 20\mevc$) ($\approx\! 2.5\,\sigma$ in each case). To improve
their momentum resolution, $D$ candidates are refitted to the nominal
$D^0$ or $D^+$ masses. To study the contribution of combinatorial
background under the $D$ peak, we use $D$ sidebands selected from a
mass window four times as large. For the study of the process $\ee
\to \jp D^* \overline{D}{}^{(*)}$ we use only the cleanest $D^{*+} \!
\to \! D^0 \pi^+$ channel. $D^{*+}$ candidates from the signal
window, selected in the interval $\pm 3\mevc$ of the nominal $D^{*+}$
mass ($\approx \! 2.5\, \sigma$), are refitted to the nominal
$D^{*+}$ mass. The $D^{*+}$ sideband region is defined by $2.016\gevc
\!  <\! M(D^0\pi^+)\!<\!2.028\gevc$. Only one $\jp D$ or one $\jp
D^{*+}$ combination per event is accepted; the combination with the
best sum of $\chi^2$ of the mass fits for \jp\ and $D^(*)$ candidates
is selected. In the $D^{(*)}$ sidebands a single candidate per event
is selected as well. The sideband is divided into windows of the same
width as the signal one, and the candidate with the smallest
difference in mass from the center of its window is chosen.


The method for reconstructing the processes \eedda\ was described
in~\cite{x3940}. In addition to the fully reconstructed \jp, only one
of the \da's is fully reconstructed (referred to below as \darec:
$\drec\!=\!D^0$ or $D^+$, $\dsrec\!=\!D^{*+}$), and the other
unreconstructed \dab\ (referred to as associated $D\equiv\datag$) in
the event is observed as a peak in the spectra of masses recoiling
against the reconstructed combination \jp\darec. The recoil mass
against the particle or combination of particles is defined as
\begin{equation}
\RM(X) = \sqrt{(E_{\rm CM}-E_{X}^{*})^2-p_{X}^{*~2}},
\end{equation}
where $E^*_{X}$ and $p_{X}^*$ are the CM energy and momentum of the
(combination of) particle(s). The $\RM(\jp\darec)$ peak around the
nominal mass of \datag\ with a typical resolution $\sim \! 30\mevc$
is used to identify the studied process. As the resolution is smaller
than $ M_{D^*} -M_D$, the method allows the contributions from the
processes \eedd, \jp\dds\ and \dsds\ to be disentangled.  The
$\RM(\jp \drec)$ and $\RM(\jp \dsrec)$ spectra in the data are shown
in Fig.~\ref{rmx} as points with error bars for the signal \darec\
windows; histograms show the scaled \darec\ sideband distributions.
The signals for the processes \eedd, \dds\ and \dsds\ are evident in
Fig.~\ref{rmx}\,a) at the $D$ and $D^*$ nominal masses and at a mass
$\sim \! 2.2\gevc$, respectively. The latter peak is shifted and
widened due to two missing pions (or photons) from $D^*$ decays.
Another excess at $\sim \! 2.45\gevc$ can be explained by the process
$\ee \! \to \! \jp D \overline{D}{}^{**}$. The processes \eedds\ and
\dsds\ are also clearly seen in Fig.~\ref{rmx}\,b) as distinct peaks
around the $D$ and $D^*$ nominal masses. We use \darec\ sidebands to
describe the combinatorial background contribution through
simultaneous likelihood fits to the \datag\ signal and sideband
spectra. The signal shapes are fixed from the Monte Carlo (MC)
simulation. The background distribution is parameterized by a
second-order polynomial function (linear function in case of \dsrec).
Only the region below $2.35\gevc$ is used because of a possible
contribution from $\ee \! \to \! \jp D \overline{D}{}^{**}$. The
signal yields (including the tail due to initial state radiation
[ISR]) and statistical significances are listed in
Table~\ref{tab:rmx}.
\begin{figure}[t]
\includegraphics[width=0.75\textwidth]{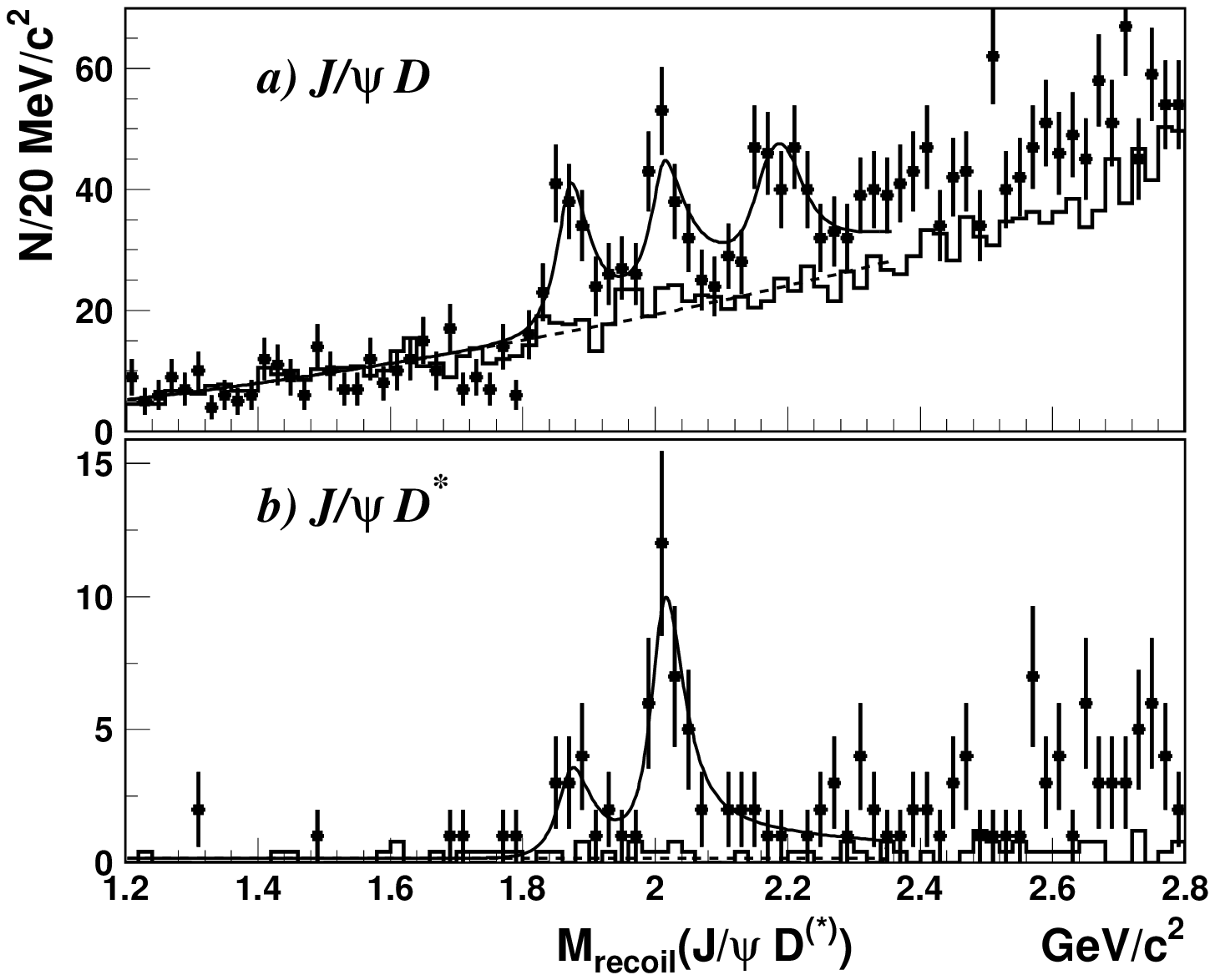}
\caption{The distributions of masses recoiling against the
  reconstructed a) $\jp D$ and b) $\jp D^*$ combinations in the
  data. The histograms show the scaled $D^{(*)}$ sideband
  distributions. The solid curves are results of the fit, the dashed
  curves are the background functions.}
\label{rmx}
\end{figure}

\begin{table}[htb]
\caption{Summary of the results of the fits to the $\RM(\jp \drec)$
  and $\RM(\jp \dsrec)$ spectra.}
\label{tab:rmx}
\begin{center}
\begin{tabular}
{l||c|c||c|c} \hline \hline & \multicolumn{2}{c||}{\jp \drec} &
\multicolumn{2}{c}{\jp \dsrec } \\ \cline{2-5}

& N & $\mathcal{N}_\sigma$ & N & $\mathcal{N}_\sigma$ \\ \hline

\eedd & \ \ $162\pm 25$ \ \ & \ \ 7.6 \ \ & --- & --- \\

\eedds & $159\pm 28$ & 6.5 & \ \ $19.0\,^{\,+6.3}_{-\,5.3}$ \ \ & \ \ 5.8 \
\ \\

\eedsds & $173\pm 32$ & 5.6 & $47.2\,^{+\,8.5}_{-\,7.8}$ & 8.4 \\

\hline \hline
\end{tabular}
\end{center}
\end{table}


We perform a study of these observed processes and search for new
charmonium states $X_{c\bar{c}}$ that can be produced via $\ee \! \to
\! \jp X_{c\bar{c}}$ followed by the decay $X_{c\bar{c}} \! \to \!
\dda$. Tagging the process \eedda\ by the requirement $|\RM (\jp
\darec) - M_{D^{(*)}}| < 70\mevc$ we thus divide each of selected
$\jp D$ or $\jp D^{*+}$ combinations into two non-overlapping
samples, each comprising $\sim \! 50\%$ of the signal events. The ISR
tail causes an inefficiency for the tagging requirement as well as
cross talk between different final states: the contribution of the
process \eedd\ (\dds) to the sample tagged as \eedds\ (\dsds) is
$\sim\! 10\%$, the reverse cross talk is only $\sim\! 1.5\%$ and
neglected. In our study we constrain $\RM(\jp \darec)$ to the \datag\
nominal mass. This improves the resolution on $\RM(\jp)$, which
corresponds to the invariant mass of the produced $D^{(*)}$ meson
pair, by a factor of $3-10$ with respect to the unconstrained value
($\sim\!30\mevc$). The $M(\dda)$ resolution varies from 2\mevc\ at
threshold to 8\mevc\ at $M(\dda)=5.0\gevc$ for all the processes
except \eedds\ with \drec\dstag. In the latter case the resolution is
worse because of the \drec\ from the $D^*$ decay ($\sim \! 10\mevc$
at $M(\dds) \! \sim \! 3.94\gevc$).

In the data the spectra of $M(\dda)$ are shown in Figs.~\ref{dd}\,a),
b), c), d) for \drec\dtag, \drec\dstag, \dsrec\dtag, \dsrec\dstag\
cases, respectively. Points with error bars correspond to the \darec\
signal windows while hatched histograms show the scaled \darec\
sideband distributions. Excesses from the signal \drec\ window over
the sideband distributions are seen around the threshold in all
figures. The reflections ($\dd \! \to \! \dds$ and $\dds \! \to \!
\dsds$) estimated using the MC simulation are shown with open
histograms. In the MC the \eedda\ processes are generated with
$M(\dda)$ spectra tuned to the data.
\begin{figure}[htb]
\includegraphics[width=0.75\textwidth]{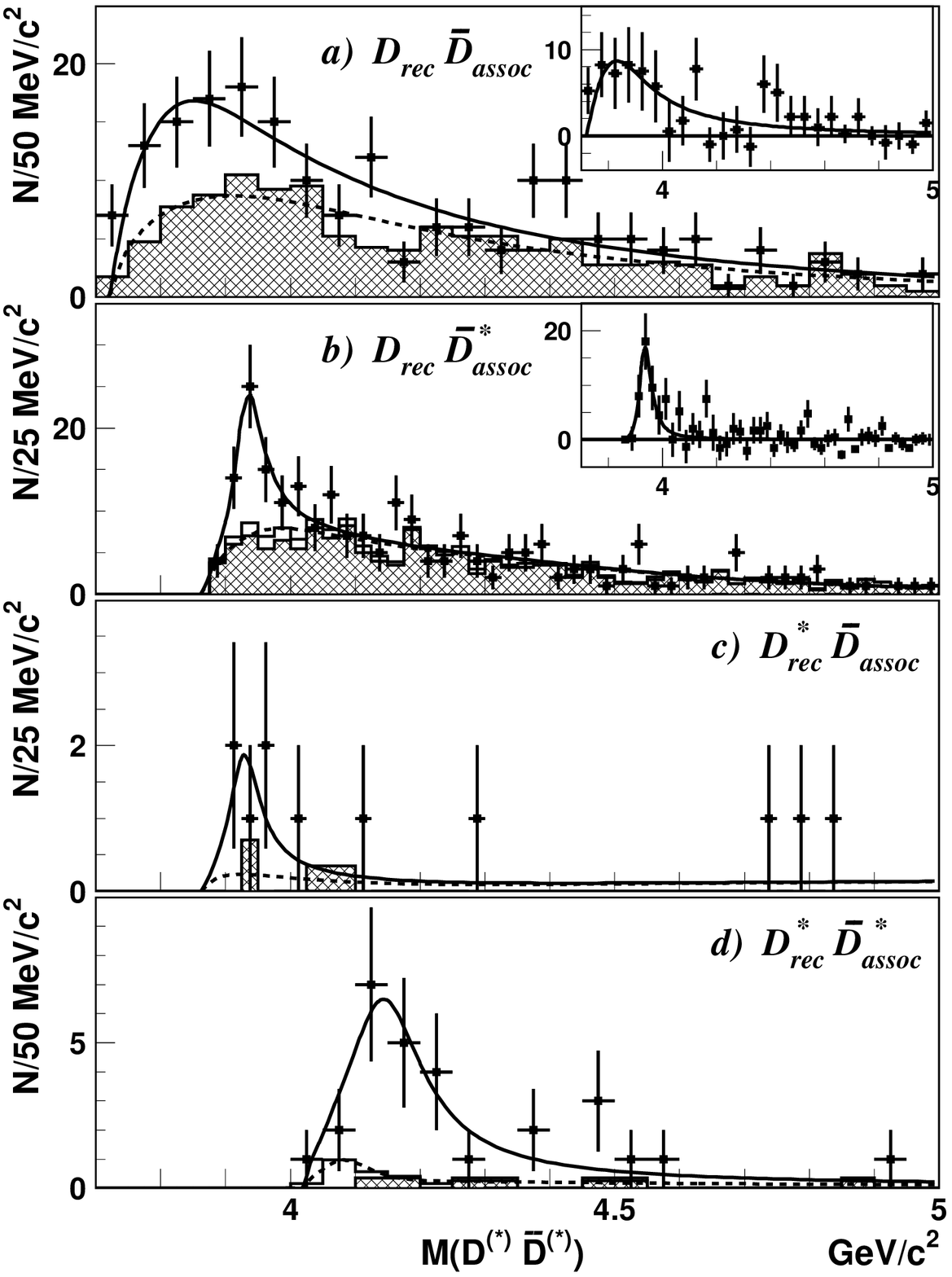}
\caption{The $M(\darec\datag)$ spectra for events tagged and
  constrained as a) \eedd, b), c) \eedds\ and d) \eedsds\ in the
  data. } \label{dd}
\end{figure}

We perform simultaneous likelihood fits to \darec\ signal and
sideband distributions to fix the combinatorial background shapes.
The accuracy of description of combinatorial backgrounds by \darec\
sidebands is validated with the MC simulation and with the data,
where the $M(\dda)$ spectra in the different sideband intervals are
found to be in good agreement with each other. The combinatorial
backgrounds are parameterized by the function $A
\sqrt{M-M_{\text{thr}}}\cdot e^{-B\cdot
  M}$, where $A$ and $B$ are free parameters, except for the case
\drec\dtag, where this shape is found to describe poorly the behavior
of the background. In the latter case we parameterize the background
by a relativistic Breit-Wigner function with a free mass, width and
amplitude. The signal functions are a sum of a relativistic $s$-wave
Breit-Wigner function and a threshold function
($\sqrt{M-M_{\text{thr}}}$) to account for possible non-resonant
production. The signal functions are convolved with the resolution
functions and multiplied by the efficiency function obtained from the
MC simulation. The reflections are taken into account in the fit.

The fitted parameters of the Breit-Wigner functions and significances
of the resonance contributions are listed in Table~\ref{tab:bw}.  We
assess the significance of each signal using $-2\ln(\mathcal{L}_0 /
\mathcal{L}_{\text{max}})$, where $\mathcal{L}_{\text{max}}$ is the
maximum likelihood returned by the fit, and $\mathcal{L}_0$ is the
likelihood with the amplitude of the Breit-Wigner function set to
zero. This quantity should be distributed as $\chi^2(n_{dof}=3)$ in
the absence of signal, as three signal parameters are free in the fit
for $\mathcal{L}_{\text{max}}$. The non-resonant contributions are
consistent with zero within $1 \, \sigma$ in all fits, except for the
case \dsrec\dtag\ (1.6$\,\sigma$ from zero). The fit results are shown
in Fig.~\ref{dd} as the solid curves; the dashed curves are the
background functions. The insets in Fig.~\ref{dd}\,a) and b) show the
background subtracted spectra with the signal functions superimposed.

\begin{table}[htb]
\caption{Summary of the signal yields, masses [\mevc], widths [\mev]
  and significances for $\ee \to \jp (\dda)_{\text{res}}$.}
\label{tab:bw}
\begin{center}
\begin{tabular}{l||c|c|c|c}
\hline  \hline

State & $N_{events}$ & $M$& $\Gamma$ &
$\mathcal{N}_\sigma$ \\ \hline

$X(3880)$(\drec\dtag) & $63\,^{+\,31}_{-\,25} $ & $\, 3878 \pm 48\,$ &
$\, 347\,^{+\,316}_{-\,143} \, $ & $\,3.8\,$ \\

$X(3940)$(\drec\dstag) & $52\,^{+\,24}_{-\,16}$ &
$3942\,^{+\,7}_{-\,6}$ & $37\,^{+\,26}_{-\,15}$ & 6.0 \\

$X(3940)$(\dsrec\dtag) & $5.2\,^{+\,3.4}_{-\,2.7}$ &
$3934\,^{+\,23}_{-\,17}$ & $57\,^{+\,62}_{-\,34}$ & 2.8 \\

$X(4160)$(\dsrec\dstag)\, & $\,23.8\,^{+\,12.3}_{-\,\phantom{1}8.0}\,$
& $4156\,^{+\,25}_{-\,20}$ & $139\,^{+\,111}_{-\,\phantom{1}61}$ & 5.5
\\ \hline \hline
\end{tabular}
\end{center}
\end{table}

A fit to $M(\dd)$ distribution finds a broad resonance near the
threshold, which is tentatively denoted as $X(3880)$, with a
statistical significance of $3.8\,\sigma$. However, the fit is not
stable under variation of background parameterization as well as
variation of the bin width. The fit with two resonances better
describes the spectrum and is more stable, but the significance of
the second resonance is lower than $3\,\sigma$. We conclude that the
observed threshold enhancement is not consistent with non-resonant
$\ee \! \to \! \jp \dd$ production, but with the present statistics
the resonant structure in this process cannot be reliably determined.
The significance of the $X(3940)$ signal found by the fit to the
$M(\drec\dstag)$ spectrum is $6.0\,\sigma$. The fitted width of
$X(3940)$
is slightly higher than that obtained in our previous
analysis~\cite{x3940}. The mass of the state is in good agreement
with the reported mass, and the signal yield scales with respect to
the previous result in proportion to the luminosity. Separate fits to
the $D_{\text{rec}}^0$ and $D_{\text{rec}}^+$ samples yield
$42\,^{+\,10}_{-\,\phantom{1}9}$ and $8\,^{+\,5}_{-\,4}$ signal
events, respectively, in good agreement with the MC expectations (40
and 12) normalized to the integrated yield assuming equal branching
fractions of $X(3940)$ decays into charged and neutral $D
\overline{D}^{*}$ pairs.  The $X(3940)$ signal is also seen in the
$M(\dsrec\dtag)$ spectrum with a significance of $2.8\,\sigma$, with
parameters in good agreement with those from the $M(\drec\dstag)$
fit. As this sample is a small subsample of the \drec\dstag\ case, we
use the latter fit for only as a cross check. The $M(\dsds)$ spectrum
demonstrates a clear broad enhancement around the threshold, which we
denote as $X(4160)$. The $X(4160)$ signal is seen above the small
combinatorial background and the $X(3940)$ reflection with a
statistical significance of $5.5\,\sigma$.

The Born cross sections for the processes $\ee \! \to \! \jp
X(3940)~[X(4160)]$ multiplied by
$\mathcal{B}_{D^{(*)}\overline{D}{}^*} \equiv \mathcal{B}(X \to
D^{(*)} \overline{D}^{*})$ are calculated from the fitted $X(3940)$
and $X(4160)$ yields with the procedure used in the previous
analysis~\cite{2cc2}. Taking into account the reconstruction
efficiencies obtained from the MC simulation, the calculated Born
cross-sections are:
\begin{eqnarray}
\sigma (\ee \! \to \! \jp X(3940)) \mathcal{B}_{D^*\overline{D}} =
(13.9^{+6.4}_{-4.1} \pm 2.2)\, \text{fb} \nonumber \\ \sigma (\ee \!
\to \! \jp X(4160)) \mathcal{B}_{D^*\overline{D}{}^*} =
(24.7^{+12.8}_{-\phantom{1}8.3} \pm 5.0) \,\text{fb}.
\end{eqnarray}

The systematic errors of the parameters and production cross sections
for $X(3940)$ and $X(4160)$ resonances are summarized in
Table~\ref{sys}. To estimate the fitting systematics we study the
difference in $X(3940)$ [$X(4160)$] parameters returned by the fit to
the Fig.~\ref{dd}\,b) and d) distributions under variation of the
signal and background parameterizations, the fit ranges and the
histogram bins as well as the resolution functions. We also vary the
definitions of the signal and sideband regions to check the stability
of the resonance parameters. Another uncertainty in the determination
of the masses is due to possible momentum scale bias. This was
estimated in the previous paper~\cite{x3940} to be smaller than
$3\mevc$. The systematic error for the cross section calculation is
dominated by the uncertainty in the \jp\ production and polarization
angular distributions. In the MC both angular distributions are
assumed to be flat and extreme cases ($1+ \cos^2{\theta}$ and
$\sin^2\theta$) are considered to estimate the systematic uncertainty
in this assumption. In the case of the $X(4160)$ another source of
the systematic uncertainty is the $D^{*+}$ polarization, which is
also taken into account by varying the $D^{*+}$ helicity angle
distribution. Other contributions come from the uncertainty in the
track and \pin\ reconstruction efficiencies; lepton and kaon
identification and in the absolute $\mathcal{B}(D^{(*)})$.

\begin{table}[]
\caption{Summary of the systematic errors in the masses ($M$ in
  \mevc), widths ($\Gamma$ in \mev) and production cross sections
  ($\sigma$ in \%) for $X(3940)$ [$X(4160)$] resonances.}
\label{sys}
\begin{center}
\begin{tabular}{l||c|c|c||c|c|c}
\hline \hline

& \multicolumn{3}{c||}{$X(3940)$} & \multicolumn{3}{c}{$X(4160)$} \\
\cline{2-7}

Source & \ \ $M$ \ \ & \ \ $\Gamma$ \ \ & \ \ $\,\sigma\,$ \ \ ~ & \ \
$M$ \ \ & \ \ $\Gamma$ \ \ & \ \ $\,\sigma\,$ \ \ \\

\hline

Fitting procedure & $\pm 4$ & $\pm 6$ & $\pm 5$ & $\pm 12$ & $\pm 18$
& $\pm 2$\\

Selection & $\pm 4$ & $\pm 5$ & $\pm 4$ & $\pm 8$ & $\pm 11$ &
$\pm 5$\\

Momentum scale & $\pm 3$ & --- & --- & $\pm 3$ & --- &
--- \\

Angular distributions & --- & --- & $\pm 12$ & --- & --- & $\pm 16$ \\

Reconstruction & --- & --- & $\pm 6$ & --- & --- & $\pm 8$ \\

Identification & --- & --- & $\pm 4$ & --- & --- & $\pm 4$ \\

$\mathcal{B}(D^{(*)})$ & --- & --- & $\pm 3$ & --- & --- & $\pm 4$ \\

\hline Total & $\pm 6$ & $\pm 8$ & $ \pm 16 $ & $ \pm 15 $ & $\pm 21$
& $\pm 20$\\

\hline \hline
\end{tabular}
\end{center}
\end{table}

In summary, we have observed the processes \eedd\ (\dds, \dsds) and
found significant enhancements in $M(\dda)$ spectra around thresholds
in all these processes. A broad enhancement in $M(\dd)$ is not
consistent with non-resonant $\ee \! \to \! \jp \dd$ production,
however the present sample is not large enough to allow the resonant
structure in this process to be determined. We have confirmed our
observation of the charmonium state, $X(3940)\! \to \! D
\overline{D}{}^*$, produced in the process $\ee \! \to \! \jp\,
X(3940)$ with a significance of $5.7\,\sigma$ including systematics.
The $X(3940)$ mass and width are $(3942\,^{+\,7}_{-\,6} \pm 6)\mevc$
and $\Gamma=(37\,^{+\,26}_{-\,15}\pm 8)$\mev.
These measurements are consistent with our published results and
supersede them. In this study we have found that the inclusive peak
in the $\RM(\jp)$ spectrum may consist of several states, thus our
previous measurement of $X(3940)$ branching fractions may be not
reliable~\cite{x3940}. We report observation of a new charmonium-like
state the $X(4160)$ in the processes $\ee \! \to \! \jp\, X(4160)$
decaying into \dsds\ with a significance of $5.1\,\sigma$, including
the systematic uncertainty of the fit. The $X(4160)$ parameters are
$M= (4156\,^{+\,25}_{-\,20} \pm 15)\mevc$ and $\Gamma =
(139\,^{+\,111}_{-\,\phantom{1}61} \pm 21)\mev$.

We thank the KEKB group for excellent operation of the accelerator,
the KEK cryogenics group for efficient solenoid operations, and the
KEK computer group and the NII for valuable computing and Super-SINET
network support.  We acknowledge support from MEXT and JSPS (Japan);
ARC and DEST (Australia); NSFC and KIP of CAS (China); DST (India);
MOEHRD, KOSEF and KRF (Korea); KBN (Poland); MES and RFAAE (Russia);
ARRS (Slovenia); SNSF (Switzerland); NSC and MOE (Taiwan); and DOE
(USA).


\begin{thebibliography} {99}

\bibitem{2cc} K.~Abe, {\it {et al.}} (Belle Collab.),
Phys. Rev. Lett. {\bf 89}, 142001 (2002).

\bibitem{2cc2} K.~Abe {\it {et al.}} (Belle Collab.), Phys. Rev. D
  {\bf 70}, 071102 (2004).

\bibitem{babar_2cc} B.~Aubert {\it {et al.}} (\bbr\ Collab.),
  Phys. Rev. D {\bf 72}, 031101 (2005).

\bibitem{cc} Charge-conjugate modes are included throughout this
  paper.

\bibitem{x3940} K.~Abe {\it et al.} (Belle Collab.),
Phys. Rev. Lett. {\bf 98}, 082001 (2007).

\bibitem{new_cc} S.K.~Choi {\it {et al.}} (Belle Collab.),
Phys. Rev. Lett. {\bf 89}, 102001 (2002);
 S.K.~Choi {\it {et al.}}  (Belle Collab.),
Phys. Rev. Lett. {\bf 91}, 262001 (2003);
 S.K.~Choi {\it {et al.}} (Belle Collab.),
Phys. Rev. Lett.  {\bf 94}, 182002 (2005);
 B.~Aubert {\it et al.} (\bbr\ Collab.),
Phys. Rev. Lett. {\bf 95}, 142001 (2005).

\bibitem{theor} T.~Barnes, S.~Godfrey and E.S.~Swanson, Phys. Rev. D
  {\bf 72}, 054026 (2005); E.J.~Eichten, K.~Lane and C.~Quigg,
  Phys. Rev. D {\bf 73}, 014014 (2006).


\bibitem{Belle} A.~Abashian {\it et al.} (Belle Collab.),
Nucl. Instr. and Meth. A {\bf 479}, 117 (2002);  Z.Natkaniec {\it et
al.} (Belle Collab.), Nucl. Instr. and Meth. A {\bf 560}, 1 (2006).

\bibitem{KEKB} S.~Kurokawa and E.~Kikutani, Nucl. Instrum. Meth. A,
  {\bf 499}, 1 (2003).


\end{thebibliography}
\end{document}